\documentclass[12pt]{article}
\usepackage[tmargin=1in,bmargin=1in,lmargin=1.25in,rmargin=1.25in]{geometry}

\usepackage{setspace}
\usepackage{csquotes}
\usepackage{amsmath}
\usepackage{amsfonts}
\usepackage{stmaryrd}
\usepackage{dsfont}
\usepackage{graphicx}
\usepackage[title]{appendix}
\usepackage[font=small,labelfont=bf]{caption}
\usepackage{lipsum}
\usepackage{mwe}
\graphicspath{ {images/} }

\title{Intrinsic Local Distances: A Mixed Solution to Weyl's Tile Argument\footnote{I thank Philip Bricker and Jeffrey Russell for very helpful guidance, feedback, and discussions. I thank the audience at my talks based on this paper in Metaphysical Mayhem at Rutgers University in 2018, and in Philosophy of Logic, Mathematics, and Physics Graduate Conference at the University of Western Ontario in 2019. Among the audience, I especially thank Cian Dorr for his helpful feedback. I'd also like to thank a referee of \textit{Synthese} for pressing me on the application of my account to relativistic settings, which helps clarify the relevance of the account.}}
\date{(Forthcoming in \textit{Synthese})}
\author{Lu Chen}
\begin{document}
	\maketitle
	
	\noindent\textbf{Abstract.} Weyl's tile argument purports to show that there are no natural distance functions in atomistic space that approximate Euclidean geometry. I advance a response to this argument that relies on a new account of distance in atomistic space, called \textit{the mixed account}, according to which \textit{local distances} are primitive and other distances are derived from them. Under this account, atomistic space can approximate Euclidean space (and continuous space in general) very well. To motivate this account as a genuine solution to Weyl's tile argument, I argue that this account is no less natural than the standard account of distance in continuous space. I also argue that the mixed account has distinctive advantages over Forrest's (1995) account in response to Weyl's tile argument, which can be considered as a restricted version of the mixed account.
	
	\vspace*{16mm}
	
	\section{Weyl's Tile Argument}

	According to \textit{the atomistic view}, space (or spacetime) is composed of extended indivisible parts---call them ``atoms." This view is motivated by both conceptual and empirical puzzles for the standard view, according to which space is composed of extensionless points (for example, see Van Bendegem 1995 and Baez 2018). However, there is a famous argument given by Weyl (1949) against it:
	
	\begin{quote}
		How should one understand the metric relations in space on the basis of this idea? If a square is built up of miniature tiles, then there are as many tiles along the diagonal as there are along the side; thus the diagonal should be equal in length to the side. (Weyl 1949, 43)
	\end{quote}

	\noindent Consider the following square region composed of $4\times 4$ atoms represented by square tiles (Figure 1):\footnote{My presentation of the argument follows Salmon (1980).}
	
	\begin{figure}[h]
	\vspace{5mm}\hspace{18mm}\includegraphics[scale=0.35]{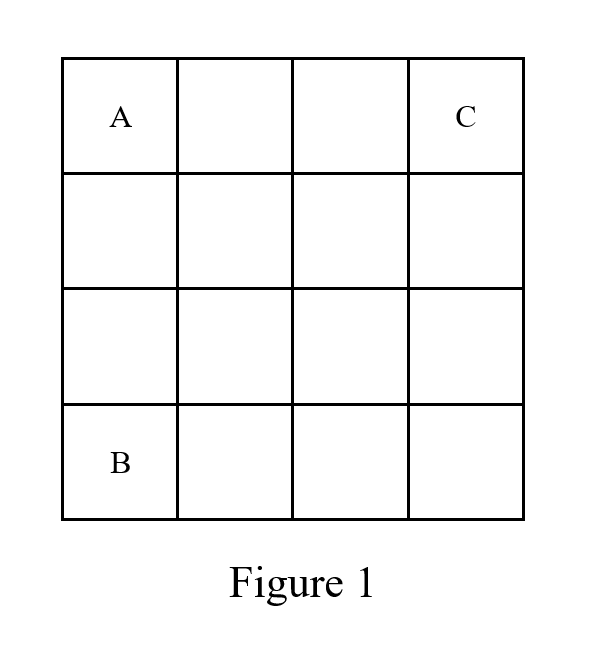}
\hspace{14mm}\includegraphics[scale=0.35]{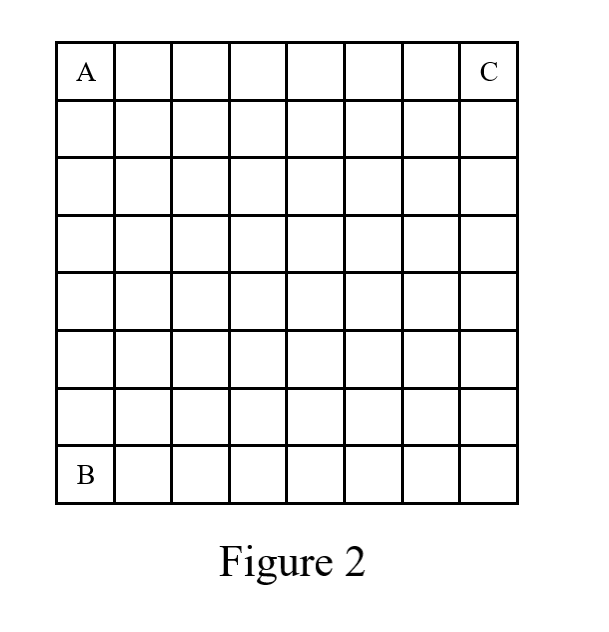}
	\end{figure}

	\noindent There are four atoms on the side $AC$. There are also four atoms on the diagonal $BC$. This, according to Weyl, implies that $AC$ and $BC$ have the same length. But if the Pythagorean theorem is approximately true, then $BC$ should be about $\sqrt{2}$ times as long as $AC$. Adding more atoms does not help. If the square is made of $8\times 8$ atoms (Figure 2), there are still as many atoms on the diagonal as on the side. So no matter how big the square region is, the ratio between the length of its side and the length of its diagonal does not approximately satisfy the Pythagorean theorem. Weyl concluded that, since the Pythagorean theorem is approximately true, our space is not atomistic.
	
	This conclusion is relevant to both philosophers and physicists. Whether space is atomistic is an active research question in physics. For example, experiments have been proposed to test the hypothesis that space is composed of ``atoms" at the Planck scale (Hogan 2012).\footnote{I put ``atoms" in quotes because it is not entirely clear what philosophical theory of spacetime we should explicate from Hogan (2012). More technically, the tested hypothesis implies that the geometry of spacetime is not commutative below the Planck level. Among other things, this means that unextended points do not exist because the coordinates of a point are necessarily commutative (e.g., in the $(x,y)$-coordinate system, for any point $(a,b)$, $ab-ba=0$). } It is a ``received wisdom" that a certain sort of discrete structure is required for reconciling quantum theory and general relativity (Maudlin 2015, 46). But if Weyl's tile argument is successful, then we can conclude that space is not atomistic without doing experiments. Due to its relevance, physicists continue to be intrigued by this argument (for example, see Crouse and Skufca 2018). 
	
	Even though Weyl's tile argument has been found ``devastating" (Van Bendegem 2019), the core assumptions that the argument relies on have not been explicitly motivated. For instance, why might we think that the length of the diagonal equals the number of the atoms on the diagonal? This, as I will explain in Section 2, amounts to a simple path-dependent account of distance in atomistic space, which fits into our best physical theory. In contrast, a perhaps equally intuitive alternative---\textit{the intrinsic account of distance}---does not have similar merits (Section 3; see also McDaniel 2007).  
	
	Making the underlying account of distance explicit is not only useful for appreciating the force of Weyl's argument, but also for opening up new options that haven't been considered so far (for current solutions, see Van Bendegem 1987, 1995, Forrest 1995). I will propose a solution to Weyl's tile argument by appealing to a new account of distance in atomistic space, called \textit{the mixed account}, according to which there are primitive distances at the small scale (Section 4). I will argue that this account is a successful reply to Weyl's argument by comparing it with the standard account of distance in standard space, which exemplifies a similar structure (Section 5). I will also argue that the mixed account has distinct advantages over Forrest's proposal (Section 6).  
	
	For simplicity, I will pretend that our actual space is Euclidean for the most part, except that in Section 5, I will focus on the standard account of distance for continuous space in general. I also briefly discuss a generalization of my account to relativistic settings in Section 4.

	\section{Path-Dependent Distance}
	
	In this section, I will identify the account of distance implicitly assumed by Weyl's tile argument and examine the rationale behind it. I will argue that there is room for rejecting this premise, and propose the conditions for a successful response to Weyl's argument. 
	
	An important step in Weyl's tile argument is to claim that the lengths of the side $AC$ and the diagonal $BC$ are both determined by the numbers of atoms they contain.  Under standard geometry, we would think that the property of length is only fundamentally instantiated by one-dimensional line segments or, more generally, a path. However, in atomistic space, there are no one-dimensional line segments or paths in the standard sense. So, how should we understand ``length" in atomistic space? A natural option is to define a new notion of ``path" in atomistic space to which the property of length fundamentally applies. The definition involves a primitive notion of \textit{adjacency} that is reflexive and symmetric.

	\begin{quote}
		\textsc{Path.} A \textit{path} from atom $a_1$ to $a_n$ is a sequence of atoms $a_1,a_2,...,a_k,...,a_n$ such that for every $k$, $a_k$ and $a_{k+1}$ are adjacent ($1\leq k\leq n-1$).
	\end{quote}
	
	\noindent Assuming that the unit of length is the length of a path containing one atom, Weyl's tile argument can be taken as relying on the following principle:

	\begin{quote}
		\textsc{Length-by-counting}. The \textit{length} of a path is equal to the number of atoms it contains.
	\end{quote}
	
	\noindent According to the standard account of distance in standard space, the distance between two points is equal to the length of a shortest path between them. This account can also apply to atomistic space:
	
	\begin{quote}
		\textsc{Path-dependent Distance}. For any two atoms $a$ and $b$, the distance between $a$ and $b$ is the length of a shortest path from $a$ to $b$.\footnote{Strictly speaking, it is more natural to think that the distance between $a$ and $b$ is the length of a shortest path from $a$ to $b$ \textit{minus one}. For example, while the length of the side $AB$ is four in Figure 1, it's more natural to think that the distance between $A$ and $B$ is three. However, for the sake of generalization in later discussions, it's better to use \textsc{Distance}.} 
	\end{quote}
	
	\noindent (Note that a shortest path from $a$ to $b$ contains the same number of atoms as that from $b$ to $a$, which implies that distance relation is indeed symmetric.) It follows that we can obtain the distance between two atoms by counting the atoms between them (Riemann 1866).
	
	In order to apply these definitions to Weyl's tile space, we need to specify which atoms count as adjacent. Weyl's tile argument amounts to endorsing the option that two atoms are adjacent \textit{iff} their representing tiles are horizontally, vertically, or diagonally adjacent. Under this stipulation, the diagonal $BC$ in Figure 2 is composed of eight atoms that are diagonally adjacent.\footnote{Another intuitive option is to assume that two atoms are adjacent \textit{iff} their representing tiles are horizontally or vertically adjacent. Under this option, the diagonal $BC$ is represented by the zigzag region along the diagonal direction (Figure 3). But this option has the same problem: the ratio of the diagonal to the side is about 2:1 rather than $\sqrt{2}:1$.
	
	\vspace{1mm}\hspace{25mm}\includegraphics[scale=0.23]{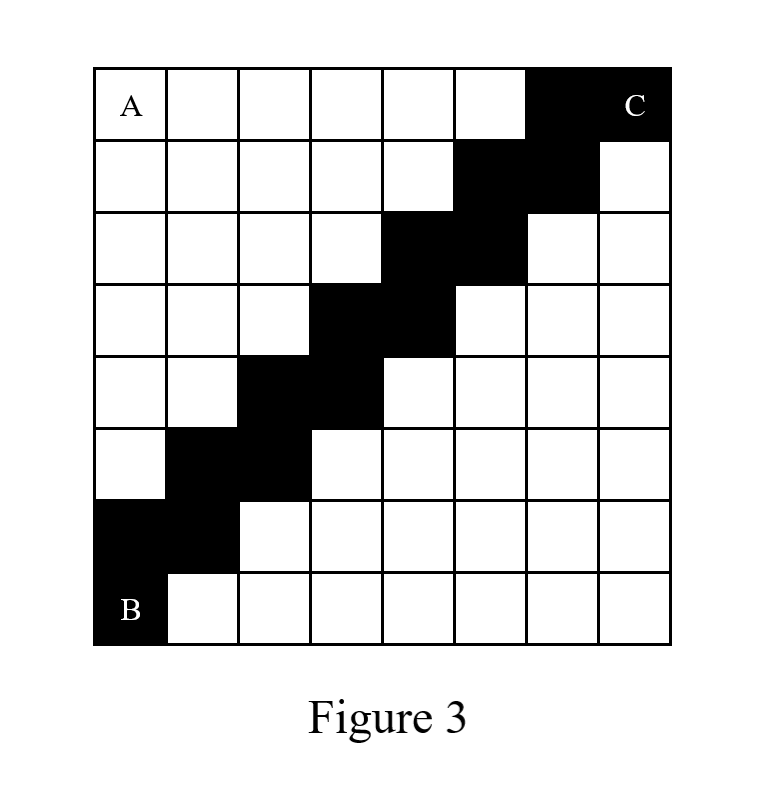}
	} 
	
	But why should we accept \textsc{Length-by-counting}? Here's one tempting thought. In standard measure theory, we have the principle of finite additivity:
	
	\begin{quote}
		\textsc{Finite additivity.} For any natural number $n$, for any $n$-dimensional region $X$, if $X$ is composed of finitely many disjoint regions $Y$s, then the measure of $X$ is the sum of the measures of $Y$s. 
	\end{quote}

	\noindent In the case of atomistic space, if we assume that every atom has a unit size, then \textsc{finite additivity} entails:
	
	\begin{quote}
		\textsc{Size-by-counting.} The measure (or size) of a region is equal to the number of atoms it contains.
	\end{quote}

 \noindent But \textsc{Size-by-counting} generally does not imply \textsc{Length-by-counting}. In standard space, a path is a one-dimensional region of space and therefore only has a one-dimensional measure. But in the case of two-dimensional atomistic space, the two-dimensional measure of a path need not be numerically equal to its one-dimensional length.\footnote{Here I am using ``dimension" in an informal (and hopefully intuitive) way that every region of $N$-dimensional atomistic space is also $N$-dimensional.
		In other words, dimensionality is an intrinsic property of an atom. But we can have alternative definitions of dimension in atomistic space, which will be briefly discussed in Section 6.} For example, imagine that an atom has a kind of shape, which is given by primitive lengths along different directions. Say an atom has a horizontal and vertical length of 1, and a diagonal length of $\sqrt{2}$, and we still assume that each atom has a unit size. In this case, the length of the diagonal $BC$ would be $4\sqrt{2}$, which is numerically unequal to the size of $BC$ (which is 4).\footnote{Note that this example does not solve Weyl's tile argument: even though the sides and the diagonal of the square region satisfy the Pythagorean theorem, the distances along other directions don't.}

Therefore, \textsc{Length-by-counting} is not a conceptual necessity for atomistic space. Forrest (1995), for example, considered it to be motivated by the consideration of theoretical simplicity and elegance. It's attractive that the metric property of space is founded on just one primitive dyadic relation of adjacency. However, simplicity and elegance should not be the sole factors for theory choice. (Besides, they are often hard to measure---a theory that is simpler and more elegant in one sense may be more complicated in other senses.) Moreover, granting that simplicity and elegance are important theoretical virtues, in order for Weyl's argument to be successful, there need to be a stronger claim, namely that there are no atomistic accounts of distance that can do as well as the standard account of distance for continuous space---for otherwise it would be unfair to conclude that our space is not atomistic but continuous. This is a claim that I will challenge in this paper. 

More explicitly, I will argue that there is an account of distance for atomistic space that meets the following ``success conditions'' and therefore solves Weyl's argument: (1) it allows atomistic space to approximate Euclidean geometry; (2) it is compatible with physics as we know it; and (3) it scores reasonably well on theoretical virtues, such as intelligibility, intuitiveness, naturalness, simplicity and so on, and in particular, it scores no worse than the standard account for continuous space. 
	
	Let me briefly address another implicit assumption in Weyl's tile argument: atoms are arranged like the regular square tiling. What if atoms are arranged very differently? For example, they may be arranged like the regular hexagonal tiles (Figure 4). We can check that the distance relations under this arrangement approximate Euclidean geometry much better than the square tiling (e.g., the ratio between the lengths of AB, AC and BC is close to what is required by the Pythagorean theorem). Nonetheless, the deviation is still large enough to be detectable at a large scale. Indeed, so far there hasn't been any clear example of tiling arrangement that approximates Euclidean geometry sufficiently well (Van Bendegem 2019).
	
	\begin{figure}

	\hspace*{45mm}\includegraphics[scale=0.15]{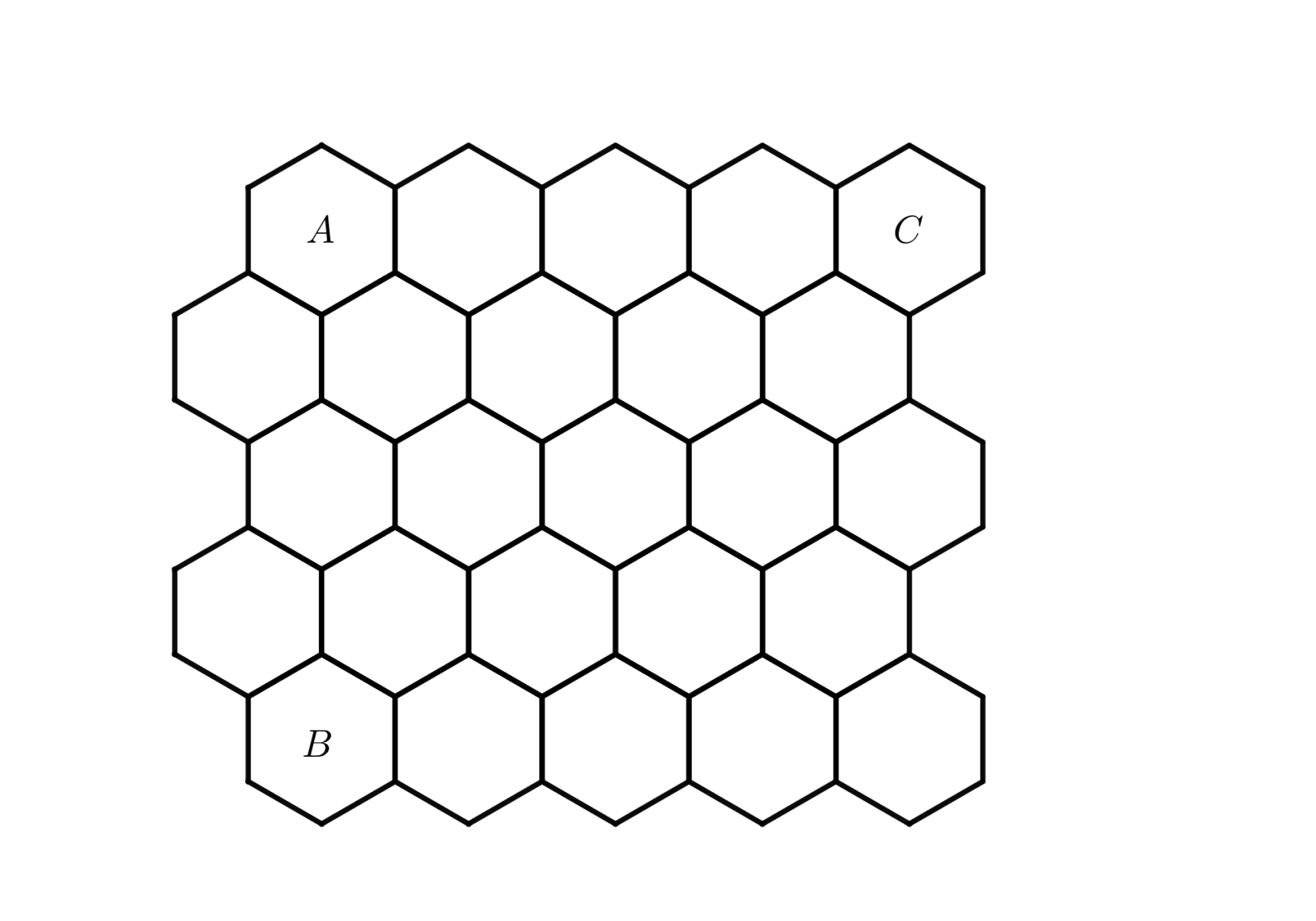} 
			\caption*{Figure 4}
	\end{figure}
	
   \noindent I don't know if there is such a tiling space, but as shown in Fritz (2013), even if there are atomistic spaces represented by some tiling arrangements that approximate Euclidean space very well at least at the large scale, those arrangements have to be very complicated and irregular.\footnote{In Fritz's formalism, atomistic space is modeled by an infinite graph composed of $\mathbb{Z}^d$-translates of a certain finite pattern---call each of those translates a ``cell." For example, in the hexagonal tile space, each cell contains just one vertex and six edges. According to Fritz, a cell must contain a very large number of edges in order for the metric of the graph to approximate Euclidean geometry closely at the large scale. This means that, if there is an atomistic space represented by a tile space that approximates Euclidean space very well at the large scale, the repeated pattern must be very complicated. I thank Fritz for clarifying the gist of Fritz (2013) in personal correspondence.} The alternative account of distance that I will propose does not depend on the existence of such a tiling arrangement and, as I will argue, solves Weyl's tile argument at least as well as a tiling-based solution. Thus the question about alternative tiling arrangements may only be of purely technical interest.

	\section{Against Intrinsic Global Distance}

	In the last section, I have been assuming the path-dependent account of distance, according to which the distance between any two atoms is equal to the length of a shortest path between them. An alternative account, \textit{the intrinsic account of distance}, says that the distance between two atoms does not depend on the path between them and indeed is intrinsic to their fusion: if we duplicate the fusion without duplicating anything else, the duplicate atoms will still have the same distance.\footnote{See McDaniel (2007) for more discussion of the view. McDaniel argued that the intrinsic account is true in some possible worlds, and in such worlds, atomistic space can approximate Euclidean distance.}
	
	Now, if the intrinsic account is true, there would be no problem assigning distances among atoms that approximate Euclidean geometry. The trick is to assign primitive distances to all pairs of atoms that match Euclidean distances.  More precisely, let each atom be represented by a pair of integers in the two-dimensional coordinate space $\mathbb{R}^2$. We assume this:
	\begin{quote}
		\textsc{Euclidean Model(I).} The distance between any two atoms $(a_1,b_1), (a_2,b_2)$ is equal to $\sqrt{(a_2-a_1)^2+(b_2-b_1)^2}$.
	\end{quote}
	\noindent Then all distance relations trivially satisfy Euclidean geometry.
	
	Have we solved Weyl's tile argument then? No, because the intrinsic account faces two objections, one empirical and one theoretical. Note that McDaniel (2007) proposed the intrinsic account as a ``solution" to Weyl's tile argument in the sense that atomistic space that satisfies \textsc{Euclidean model(I)} is metaphysically \textit{possible}. But the focus here is whether such a space is a live candidate for the structure of \textit{actual} spacetime. The answer is no, because such a space is incompatible with actual physics.  According to the theory of general relativity, the metric of spacetime is determined by the distribution of mass-energy under Einstein's field equations. Roughly, the curvature of spacetime at a point is proportional to the density of mass-energy near that point, which means that the presence of a massive body would distort the paths nearby. Furthermore, our physics is \textit{local}: there is no instantaneous action at a distance. Now, consider two spacetime points far apart. If there occurs a massive body between them, then according to general relativity, their distance will be different. But the fusion of the two points presumably does not go through any intrinsic change, especially if both points are far away from the massive body and couldn't be affected immediately. Thus the massive body changes the distance between them only by changing the length of the shortest path between them. This means that the distance between them cannot be intrinsic to them. So, the intrinsic account is false for actual space. Insofar as we want atomistic space to be a candidate for our actual space, the intrinsic account does not help. 
	
	Apart from actual physics, there is also a theoretical consideration against the intrinsic account. Maudlin (2007), among others, has argued that if distances are all primitive, then we need to posit \textsc{triangle inequality} as an axiom, which says that for any ``points" $a,b,c$, the distance between $a$ and $b$ plus the distance between $b$ and $c$ must be at least as great as the distance between $a$ and $c$.\footnote{In a general context, I use ``point" to simply refer to an ultimate part of an arbitrary space.} But if we define distance as the length of a shortest path, then \textsc{triangle inequality} automatically follows. Suppose \textsc{triangle inequality} is false: there are three points $a,b,c$ such that the length of the shortest path from $a$ to $c$ is longer than the sum of the length of the shortest path from $a$ to $b$ and that from $b$ to $c$. However, the path from $a$ to $b$ connected with the path from $b$ to $c$ \textit{just is} a path from $a$ to $c$, the length of which is equal to the sum of the two connected paths.\footnote{Here, ``connected" is used in the sense that a path $a_1,...,a_k$ can be connected with a path $a_k,...,a_n$ to form a single path $a_1,...,a_n$ ($1\leq k\leq n$).} Then, this path would be shorter than the shortest path from $a$ to $c$! Contradiction. So, a shortest path from $a$ to $c$ cannot be longer than the sum of the length of the path from $a$ to $b$ and the length of the path from $b$ to $c$. Moreover, it seems that the path-dependent conception of distance is not only sufficient but also necessary for fully justifying \textsc{triangle inequality}:  without thinking in terms of paths, it is mysterious why this axiom should hold for distance at all.\footnote{A \textit{semimetric} is a generalized distance function that does not satisfy \textsc{triangle inequality}. Under the intrinsic account, it is hard to see why a space cannot have a semimetric.} So the path-dependent account is not only simpler on this regard but also more perspicuous.
	
	Nonetheless, the above arguments only show that not every distance is intrinsic and, in particular, that global distances are not intrinsic. They do not show that no distance can be intrinsic or that the notion of intrinsic distance is unintelligible. Indeed, I will now propose an account of distance in response to Weyl's tile argument that also relies on primitive intrinsic distances.
	
	\section{Primitive Local Distance}
	
In this section, I will propose an alternative account of distance, which I call \textit{the mixed account}. According to this account, we can assign primitive distances not to all pairs of atoms but to atoms in a ``local neighborhood."  I will argue that this allows atomistic space to be approximately Euclidean and is not subject to the previous objections to the intrinsic account. 
	
Unlike the path-dependent account, we do not posit the primitive notion of adjacency. The only primitive notion we have is \textit{proto-distance}, denoted by $\textbf{d}$, which is partially defined over pairs of atoms and satisfies some standard axioms for distances ($p,q$ range over atoms):
	
	\begin{quote}
		\textsc{Nonnegativity.} $\textbf{d}(p,q)\geq 0$ if $\textbf{d}(p,q)$ is defined.
		
		\textsc{Symmetry.} $\textbf{d}(p,q)=\textbf{d}(q,p)$ if $\textbf{d}(q,p)$ is defined.
		
		\textsc{Nonsingularity.} $\textbf{d}(p,q)=0$ iff $p=q$.

	\end{quote}
	
	\noindent These proto-distances determine all the metric properties of space. For any atom, an atom that bears a primitive distance to it is a \textit{neighbor} of it, and the set of all its neighbors is its \textit{(local) neighborhood}. For the spaces we are interested in, all primitive distances are bounded by a finite number, which means that a local neighborhood is also bounded by a finite region. (Note that this requirement rules out the model under the intrinsic account of distance discussed in the last section.) Moreover, for atomistic space, we require that for any atom $a$ and any real number $r$, there are only finitely many atoms that bear primitive distances to $a$ that are less than $r$.\footnote{This condition is violated in some approaches to discrete spacetime, such as that of Crouse and Skufca (2018). According to Crouse and Skufca, a particle can jump in any direction as long as the minimal length of a step is a constant number $\chi$. This allows \textit{every} point in continuous space to be a potential position of a particle. So it may be more natural to consider their approach to be about a discrete \textit{dynamics} rather than a discrete spacetime.}  This entails that primitive distances between distinct atoms have a lower bound. Intuitively, for any finite region, there are only finitely many atoms in it.

	Next, we define the notion of a path in terms of neighbors:
	
	\begin{quote}
		\textsc{Path*}. A \textit{path} from $a_1$ to $a_n$ is a sequence of atoms $a_1,a_2,...,a_{n-1}$ such that for any $k$ with $1\leq k\leq n-1$, $a_k$ and $a_{k+1}$ are neighbors. 
	\end{quote}

	\noindent The length of a path is obtained by adding up the proto-distances along the path.

	\begin{quote}
		\textsc{Path-length.} If the sequence of atoms $a_1, a_2,...,a_{n-1}$ is a path from $a_1$ to $a_n$, then the \textit{length} of the path is equal to $\textbf{d}(a_1,a_2)+\textbf{d}(a_2,a_3)+...+\textbf{d}(a_{n-1},a_n).$   
		
	\end{quote}
	
	\noindent Just like any path-dependent account, the \textit{distance} (denoted by ``$d$") between any two atoms is the length of a shortest path from one to the other.\footnote{The construction of distance from proto-distance is closely related to the definition of geodesic distance in a weighted graph in graph theory, and to the construction of metric from semi-metric or quasi-metric (for example, see Harary 1969, Paluszy\'nski and Stempak 2009). 
		
		In more general settings, especially for continuous space, it is standard to define the distance between two points as the \textit{infimum} of the lengths of paths between them, since a shortest path between them may not exist. However, this definition coincides with my definition in the case of atomistic space due to the requirement that for any atom $a$ and any real number $r$, there are only finitely many atoms $x$ with $\textbf{d}(a,x)< r$.} Note that when proto-distances satisfy \textsc{triangle inequality}, namely $\textbf{d}(p,q)\leq \textbf{d}(p,r)+\textbf{d}(r,q)$, they are genuine distances under this account. In this case I'll call the proto-distance ``primitive distance.''

	I claim that, under this account of distance, we can find an atomistic space that approximates Euclidean space as closely as we want at all scales (see Appendix A for the proof).
	
	\begin{quote}
		\textsc{Euclidean Approximation.} Under the mixed account, for a Euclidean space of any dimension, there is an atomistic space that approximates it sufficiently well.
	\end{quote}
	
	\noindent For instance, let each atom be represented by a pair of integers. The following model approximates Euclidean space at all scales if the number $M$ is sufficiently large:
	
	\begin{quote}
		\textsc{Euclidean mixed model}. For any two atoms $a=(x_1, y_1)$ and $b=(x_2,y_2)$, the primitive distance $\textbf{d}(a,b)=\sqrt{(x_2-x_1)^2+(y_2-y_1)^2}$ if $(x_2-x_1)^2+(y_2-y_1)^2\leq M^2$; otherwise $\textbf{d}(a,b)$ is undefined.
	\end{quote}
	\noindent In this model, distances within a local neighborhood are exactly Euclidean. For two atoms that are far apart, their distances will generally differ from the corresponding Euclidean distance. But the difference can get as small as we want if we choose a sufficiently large $M$. To illustrate, consider the following region (Figure 5). 
	
	\begin{figure}[h]
				\vspace*{5mm}\hspace*{32mm}\includegraphics[scale=0.43]{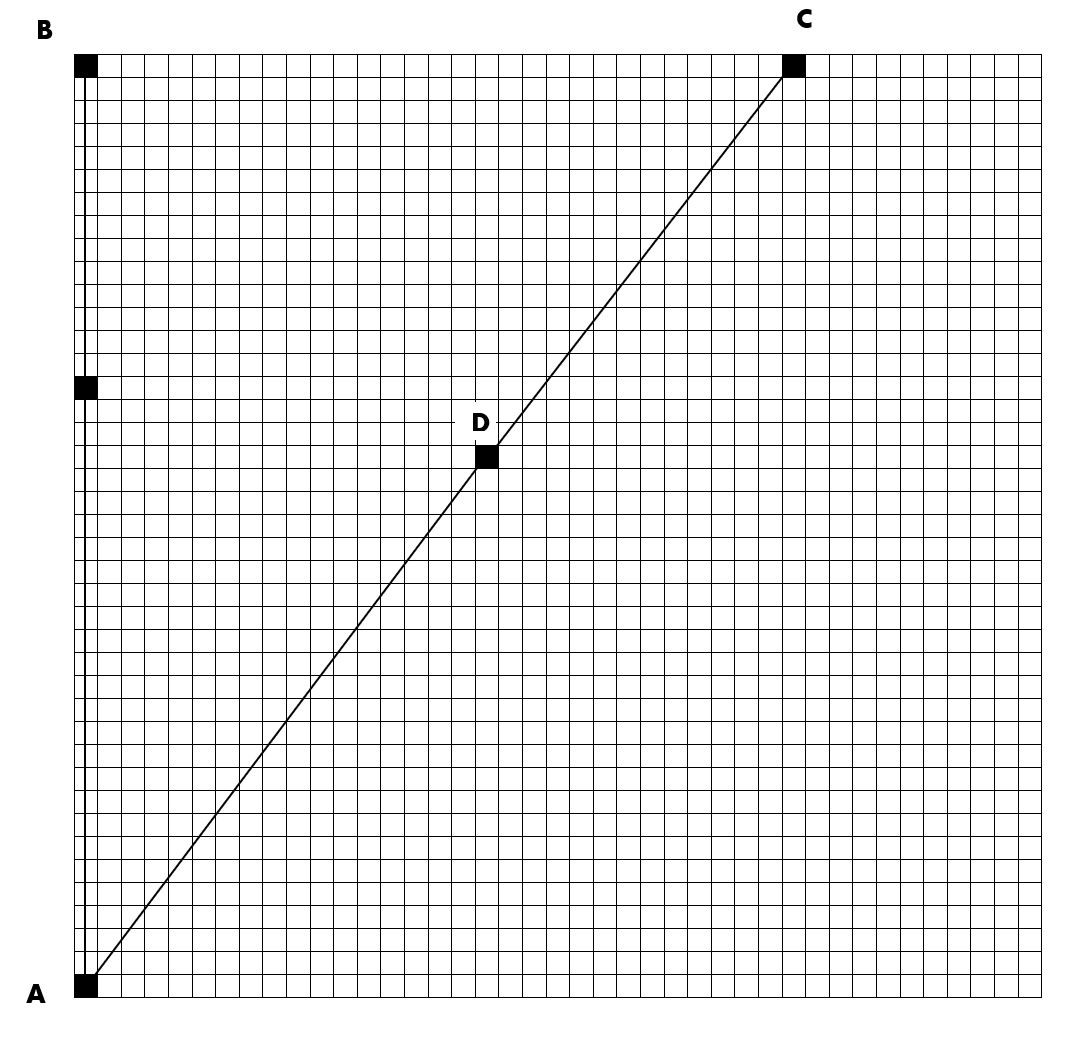} 
				\caption*{Figure 5}
	\end{figure}

	\noindent	Consider the atoms $A,B,C, D$: $A=(0,0), B=(40,0),C=(40,30),D=(17,23) $. Let $M=30$. Then $A,C$ are path-connected through $D$. This means that $d(A,C)\leq \textbf{d}(A,D)+\textbf{d}(D,B)\approx 50.01$ (and it's clear that $d(A,C)\geq 50$). It's easy to get $d(A,B)=40$ and $d(B,C)=30$. Thus, the distance relations between $A,B,C$ are very close to satisfying the Pythagorean theorem.

	The mixed account does not face the difficulties that the intrinsic account of distance has. Recall that Maudlin has objected to the intrinsic account for the reason that it needs an additional axiom of \textsc{triangle inequality}. In the mixed account, we do not need to posit this axiom. Since the distance between any two atoms is defined to be the length of a shortest path, \textsc{triangle inequality} automatically follows. Note that proto-distances need not be distances. For example, if the proto-distance between $a$ and $c$ is longer than the sum of the proto-distances between $a$ and $b$ and between $b$ and $c$, then the sequence of atoms $a,b,c$ is a shorter path from $a$ to $c$ than the sequence of atoms $a,c$. Thus, the distance between $a$ and $c$ is not the proto-distance between them. In this case, the proto-distance between $a$ and $c$ does not play any role in determining other distances either. Then, insofar as physics only involves distances and insofar as the goal is to recover physics, there is no need to posit such a proto-distance. That is, we generally do not need models where proto-distances do not satisfy \textsc{triangle inequality}.

	Since the mixed account allows large-scale distances to be path-dependent, it is compatible with our actual physics as far as we know it. For two atoms that are sufficiently far apart, their distance is not intrinsic to their fusion but depends on other atoms that compose the shortest path between them. So, when the presence of a massive body curves the shortest path between them, their distance will change accordingly. Note that the same empirical problem for large-scale primitive distances could, in principle, arise for small-scale primitive distances. So it is possible for the account to be disconfirmed by a new development in physics, supposing we can find a way to (indirectly) observe those small-scale distances and how they can be affected.
	
		Note that the mixed account can be extended to relativistic settings relatively straightforwardly, though I won't go into much detail. Here's a sketch of a possible approach. Instead of symmetric primitive distances, we may posit directed (thus antisymmetric) \textit{time-like primitive distances}, which will be sufficient for determining the metric structure of relativistic spacetime. For any atom $a$, we call atom $b$ a \textit{future neighbor} of $a$ if there is a directed primitive time-like distance from $a$ to $b$. A \textit{time-like path} is a sequence of atoms with each one preceding a future neighbor of it, and the length of a path is obtained by summing up the primitive time-like distances along the path. For any two atoms that are connected by a time-like path, the \textit{time-like distance} between them is equal to the length of a \textit{longest} path between them (a time-like distance is the \textit{maximal} time spent on traveling from one spatiotemporal atom to another).   We can derive the metric structure of spacetime from time-like paths through standard radar methods (for example, see Rosser 1992, Perlick 2007).  Here's a model for Minkowski spacetime under this extended mixed account. Let each atom be represented by a quadruple of integers $\langle t, x,y,z\rangle$.  For any two atoms $a=\langle t_1,x_1,y_1,z_1\rangle,b=\langle t_2, x_2, y_2,z_2\rangle$, if $g=(t_2-t_1)^2-(x_2-x_1)^2-(y_2-y_1)^2-(z_2-z_1)^2\geq 0$  and $g\leq M^2$, and if $t_2\geq t_1$, then the directed time-like primitive distance $\overrightarrow{\textbf{d}}(a,b)=\sqrt{g}$. Like in the case of Euclidean space, if $M$ is sufficiently large, then this atomistic model approximates Minkowski spacetime.
		
		It might be worth mentioning that this ``time-first'' approach to the discrete analogue of relativistic spacetime bears some similarity to causal set theory (Sorkin 1990). One main difference is that my approach allows for the additional structure of primitive distances rather than just a partial ordering between atoms. This additional structure may allow us to circumvent certain technical difficulties that have arisen in causal set theory.

	To take stock, the mixed account allows for an atomistic model that approximates Euclidean space (or other continuous spaces) well enough and does not face the objections the intrinsic account faces. It remains to be seen whether the mixed account scores reasonably well on other theoretic virtues. While the notion of primitive distance and of path-dependent distance are sufficiently intelligible by themselves, the mixture of the two notions may seem unnatural. To defend this account further, I will turn to a comparison between the mixed account and the standard account for continuous space.

	\section{``Local Distances" in Continuous Space}

	The mixed account might strike one as unnatural or overcomplicated because it involves two concepts of distance and allows the geometry of atomistic space to be determined by a vast number of varied primitive distances. This might lead one to uphold Weyl's conclusion that our space is not atomistic after all. Against this, I will argue that the standard account of distance from differential geometry has a similarly mixed form.  Under the standard account, as I will explain, we start with local metrics, which are analogous to primitive distances, and similarly obtain distances by ``adding up" those ``primitive distances" (though technically, it is integration rather than addition).  Note that, in arguing for this, I will shift attention from Euclidean space to generally non-Euclidean continuous space---after all, our actual space is, strictly speaking, non-Euclidean.\footnote{The mixed account can accommodate curved space as well. I will not go into details here, but one can refer to Forrest (1995, 334-40), in which Forrest explained how an atomistic model can approximate curved space once we have a model that approximates Euclidean space.} In addition, I will compare the mixed account with the standard account on how they fit into Lewisian metaphysical framework and argue for an advantage of the mixed account on this aspect.

	According to the standard account, a \textit{path} in a space is a continuous function from a unit interval to that space. We can take the unit interval in question as a unit interval of time. Then a path can be considered as the trajectory of a point-sized object in a unit interval of time. We will focus on a path that is smooth and does not intersect itself. At every point on the path, we can define a \textit{tangent vector} to be the derivative of the path at that point, which indicates the ``velocity" of the path at that point (Figure 6). 

\begin{figure}[h]
		\vspace{3mm}\hspace*{35mm}\includegraphics[scale=0.8]{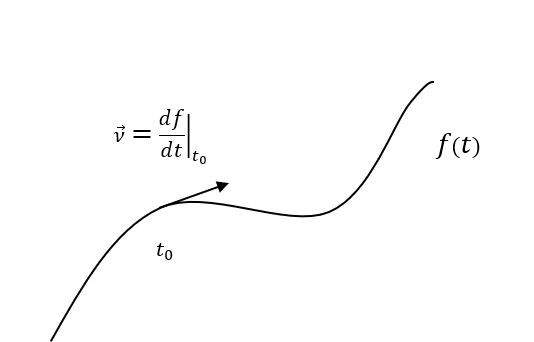}
		\caption*{Figure 6}
\end{figure}

	\noindent A \textit{metric tensor} at a point assigns a \textit{length} to each tangent vector. Heuristically, it may be helpful to think of the length of a tangent vector at a point to be the infinitesimal distance from the point along the direction of the vector divided by an infinitesimal time---though strictly speaking there are no infinitesimals in standard analysis.\footnote{For instance, in two-dimensional Euclidean space (or any flat two-dimensional Riemannian manifold), the length of a tangent vector expressed by $(\frac{dx}{dt}, \frac{dy}{dt})$ is $\sqrt{(\frac{dx}{dt})^2+(\frac{dy}{dt})^2}$.} The length of a path is obtained by integrating the lengths of the tangent vectors along the path---or informally, by adding up those ``infinitesimal distances" over the unit time interval.\footnote{More formally, consider a path in two-dimensional Euclidean space. Let $g$ be a metric tensor and $T$ range over tangent vectors along a path. Then the length of that path is $\int\sqrt{g(T,T)}dt$.}

	\begin{quote}
		\textsc{Riemannian Conception}.  The length of a path is equal to the path integral of the lengths of the tangent vectors along the path. (Riemann 1866)
	\end{quote} 
	
	\noindent The distance between any two points is the length of an extremal path between them (e.g., a longest path in Minkowski space). 
	
	Thus, like the mixed account, the standard account has a mixture of two levels. The lengths of tangent vectors  assigned by metric tensors are analogous to primitive local distances in the mixed account, and obtaining distances through integrating those lengths are analogous to adding up the primitive distances. Since we accept this mixed form in the standard account as unproblematic, we should not object to the mixed account on this ground---it is just as natural as in the standard account. (Note that ``primitive" only means ``geometrically primitive," so the primitiveness of metric tensors is compatible with their being determined by the mass-energy distribution in a dynamical theory. Analogously, we can also allow primitive distances in the mixed account to be determined in this way.) 
	
	 What's more, the primitive ``distances" in the standard account are no sparser than in the mixed account: a metric assigns a length to every tangent vector at each point, and there are infinitely many tangent vectors at every point. Since we do not know any simple foundation that can determine a geometry that is generally non-Euclidean, it seems unfair to charge the mixed account of unnaturalness and overcomplexity, at least not without further arguments.

Let's turn to another possible metaphysical reason to favor the standard account over the mixed account: local metrics are local properties and should be preferred to primitive binary relations. The idea can be captured by Lewis's well-known statement of Humean supervenience: ``all there is to the world is a vast mosaic of local matters of fact, just one little thing and then another." (Lewis, 1986: ix) In this picture, the fundamental properties or relations of the world are intrinsic properties of point-sized objects together with spatiotemporal relations (ix-x). Although Lewis himself did not consider fundamental metric features to be intrinsic properties of spatiotemporal points, it would be attractive to have \textit{a Humean geometry} which says so.
	
	The problem of invoking Humean supervenience here, however, is that it's hard to see how we could fit the standard account into the ``Humean mosaic": metric tensors and tangent vectors aren't obviously anything like qualities distributed over spacetime, and there isn't an obvious way that one can reduce the former to the latter. Of course, we can still give a try. To start fitting the standard account into a Humean geometry, let's suppose a tangent vector represents a property of a point. Here's one way of putting it: a tangent vector $v$ at a point $p$ represents $p$'s property of being such that there is a path $f$ passing through $p$ with the ``velocity" $v$. Next, we need such a property to be intrinsic to the point. But this conflicts with the standard account under the Lewisian-Humean framework. Hume famously denied necessary connections between distinct entities, which, according to Lewis, implies that it should be possible to have a perfect duplicate of an entity regardless of how the rest of the world is. Moreover, a property is intrinsic iff it never differs between perfect duplicates. So, if a property is intrinsic to a point, it should be possible to have a duplicate of that point which has that property even if the duplicate exists all by itself. However, according to the standard account, if we simply have an isolated point, then there is only ``a null vector" (a vector with length zero) at the point. So if we ``delete" the surroundings of a point in a continuous space, then the originally non-null tangent vectors at the point would become a null vector. Thus it seems natural to think that a tangent vector or the property it represents is not intrinsic to a point. This picture, then, doesn't exactly depict a Humean geometry.

	On the other hand, we do have a consideration from the Humean-Lewisian framework in favor of the mixed account over the standard account. According to Lewis, duplicates are defined in terms of \textit{perfectly natural} properties and relations. 
	
	\begin{quote}
		\textsc{Duplicate.} Two possibilia $X,Y$ are \textit{duplicates} iff there is a one-to-one correspondence between parts of $X$ and parts of $Y$ that preserves all perfectly natural properties and relations.
	\end{quote}
	
	\noindent Meanwhile, intrinsic properties and relations are defined in terms of duplicates: they are properties or relations that do not differ between duplicates.\footnote{The definition of ``intrinsic'' is adapted from Lewis, although he did not apply the term to relations (see Bricker 1993). } It follows that all perfectly natural properties and relations are intrinsic. But it is hard to see how we can fit the standard account into this framework. We have already seen that, according to the standard account, the properties represented by tangent vectors are not intrinsic to the points. But these properties are presumably perfectly natural. We would then have perfectly natural and extrinsic properties, which is incompatible with the Humean-Lewisian framework. There have been attempts of resolving this tension by revising the standard account or the Lewisian framework, which I will not discuss here.\footnote{For example, Weatherson (2006) argued that we should define duplicates in terms of fundamental properties and relations in a way that weeds out neighborhood-dependent aspects. Bricker (1993) suggested that local metrics are distances in infinitesimal neighborhoods of points.}  But it's helpful to note that the mixed account does not face this difficulty. According to the mixed account, primitive distances are perfectly natural: the fusion of atoms $a,b$ and the fusion of atoms $a',b'$ are geometric duplicates iff $a',b'$ have the same primitive distance as $a,b$. Primitive distances are also intrinsic to their relata: if you duplicate the fusion of $a$ and $b$ without duplicating anything else, their primitive distance remains the same.  So the Lewisian framework is directly applicable to the mixed account, and this may be considered an attractive feature of the account.

	\section{Forrest's Proposal}
	
	In this section, I will compare the mixed account with Forrest's (1995) solution to Weyl's tile argument.\footnote{Van Bendegem (1987, 1995) also proposed solutions to Weyl's tile argument. I consider his later proposal as a restricted version of Forrest's account. We can have a one-to-one correspondence between \textit{points} (a technical notion) in Bendegem's model and atoms in Forrest's model that preserves distance. But Forrest's account allows models that are incompatible with Bendegem's account.}  Forrest's account, as I shall argue, is a restricted version of the mixed account. While the two accounts are not mutually exclusive, the mixed account has the advantage of allowing potentially better models for actual space that are incompatible with Forrest's account.

	Like the path-dependent account in Section 2, Forrest posited exactly one fundamental dyadic relation between atoms, \textit{adjacency}, which is symmetric and irreflexive.\footnote{I change some of Forrest's terminology to align with mine. He calls atomistic space ``discrete space" and atoms ``points."} The distance between any two atoms is equal to the least number of atoms in a ``chain of adjacency" between them. What's new about Forrest's account is that, unlike what we have seen before,  two adjacent atoms do not need to be represented by two square tiles that are directly next to each other but can also be represented by tiles that are far apart. In this sense, the notion of ``adjacency" is analogous to the notion of ``neighbors" in the mixed account. Let each atom be represented by a pair of integers. Let $m$ be a parameter of atomistic space. To have a model that approximates Euclidean geometry, Forrest stipulated atoms to have the following adjacency relations:
	
	\begin{quote}
		\textsc{Forrest's model}. Two atoms $(x_1, y_1)$ and $(x_2,y_2)$ are \textit{adjacent} iff $(x_2-x_1)^2+(y_2-y_1)^2\leq m^2$.
	\end{quote}
	
	\noindent When $m$ is sufficiently large, this model approximates two-dimensional Euclidean space very well at the large scale.\footnote{For the proof, see Forrest (1995, 344-6).}

	Forrest's account is a restricted version of the mixed account, because every model under Forrest's account is isomorphic to a model under the mixed account, but not \textit{vice versa}. Since in Forrest's account, the distance between two adjacent atoms is one, \textsc{Forrest's model} is isomorphic to the following model under the mixed account:
	
	\begin{quote}
		\textsc{Forrest Mixed Model}. For any two atoms $a=(x_1, y_1)$ and $b=(x_2,y_2)$, the primitive distance $\textbf{d}(a,b)=1$ if $(x_2-x_1)^2+(y_2-y_1)^2\leq m^2$; otherwise $\textbf{d}(a,b)$ is undefined.
	\end{quote}
	
	\noindent This means that Forrest's solution is also available under the mixed account. 
	
	Since \textsc{Forrest's model} approximates Euclidean geometry at the large scale, and since Forrest's account seems simpler than the mixed account, one may think that we do not need to go for the mixed account and should stick with Forrest's account. One problem for Forest's account may be that it is counterintuitive to assign the tiles that are ``far apart" the same distance as atoms represented by tiles next to each other. But it's hard to press this issue further without a notion of ``far apart" independent from distance. More importantly, the mixed account allows for models like \textsc{Euclidean mixed model} (Section 4) that have distinctive advantages over \textsc{Forrest's model}, and we should not rule out those models as candidates for our actual space. 
	
	One difference between \textsc{Euclidean mixed model} and \textsc{Forrest's model} is that the former approximates Euclidean geometry at the local level while the latter does not. In \textsc{Forrest's model}, when $m$ is large, there are a large number of atoms that are equidistant from each other. There is no way to embed these atoms in Euclidean space that preserves their distances approximately. This could be a disadvantage of \textsc{Forrest's model} because our space may turn out to be locally Euclidean (or have a richer local geometry than what Forrest's account allows for). But one may resist this answer by arguing that if the local level is sufficiently small, that is, smaller than any observable distance, then the local geometry of the model cannot be disconfirmed by our empirical considerations.  For instance, suppose the Planck length ($10^{-35}$ meter) is the smallest physically meaningful unit. If we assume that the local level is smaller than the Planck scale, then the model will approximate Euclidean geometry above the Planck scale. Let atoms be represented by a grid of points in a scaled Euclidean plane such that the nearest points have a Euclidean distance of $10^{-65}$ meter. Furthermore, let two atoms be adjacent if their representative points have a Euclidean distance smaller than or equal to $10^{-35}$ meters. This means that the parameter $m$ in \textsc{Forrest's model} equals $10^{30}$, which guarantees that, at scales larger than the Planck scale, distances are approximately Euclidean.\footnote{The parameter $m=10^{30}$ is a number given by Forrest to ensure the model to approximate Euclidean geometry at the large scale (Forrest 1995, 333).}
	
	However, even if \textsc{Forrest's model} can be made compatible with any empirical observations of our space, this compatibility does not come without costs. First of all, under the above configuration, the vast majority of local distances in \textsc{Forrest's model} would not play any role in physical theories. Every distance that is a whole number times the Planck length is determined by the length of a path consisting of  $a_1,a_2,...,a_n$ such that for any $i=1,...,n-1$, the distance between $a_i$ and $a_{i+1}$ is the Planck length. In this case, for any two atoms that are represented by points with their Euclidean distance smaller than $10^{-35}$ meters, the distance between them does not play any role in determining the geometry at physically meaningful scales. These distances are extraneous to physical theories, and there are a lot of them: for every distance that is physically meaningful, there are about $1/2\cdot 10^{30}!$ distances that are not.

	Apart from the problem of extraneousness, by making the local level ``sub-physical," we would lose the empirical motivation for atomistic space. One motivation for atomistic space is that quantum theory and general relativity are incompatible below the Planck scale, so some physicists take the Planck length to be an indivisible unit of length.\footnote{For example, see 't Hooft (2016).} Thus the models in which atoms are significantly smaller than the Planck scale seem unhelpful for such physical considerations. This would make atomistic models less motivated.
	
	In contrast,  under the mixed account, we do not have to assume the local level to be smaller than the Planck scale. For example, let the shortest primitive distance be the Planck length. What would the longest primitive distance be in order to accommodate our current observations? The relative accuracy of a diffraction measurement, one of the best measurements for small distances, is about $\pm 1.6\times 10^{-9}$ (NIST, n.d.). Assuming we want space to approximate Euclidean geometry with a distortion smaller than this margin of error, a local neighborhood will have to have a diameter of $10^{-26}$ meter.\footnote{As shown in Appendix A, in order for the atomistic model to approximate Euclidean space, the longest primitive distance needs to be about as large as the shortest primitive distance divided by the permitted distortion (as expressed by ``$M>3r/\delta$" in the appendix).} This is a scale at which physics does not dispense with geometry. Thus unlike \textsc{Forrest's model}, local distances in this model are not extraneous. Such a model stays relevant to our empirical interest in an atomistic theory of space.
	
	The local geometry of \textsc{Euclidean mixed model} has another potentially attractive feature that \textsc{Forrest's model} lacks: it has a low dimension. According to Forrest, the dimension of a space is determined by the largest number of ``points" that are equidistant from each other. In \textsc{forrest's model}, the largest number of atoms that are equidistant from each other depends on the distance in question. As a result, Forrest proposed a scale-relative definition of dimensions for atomistic space:
	
	\begin{quote}
		\textsc{Dimension.} A space is $N$-\textit{dimensional relative to distance $D$} iff there are at most $N+1$ atoms that bear distance $D$ to each other.  
	\end{quote}
	
	\noindent According to this definition, \textsc{Forrest's model} has a very high dimension relative to the unit distance, since there are a vast number of atoms that are of a unit distance from each other. For example, with the parameter $m=10^{30}$, there are about $10^{60}$ atoms that bear a unit distance from each other, which means that the model is about $10^{60}$-dimensional at the local level. In contrast, \textsc{Euclidean mixed model} is exactly two-dimensional at the local level.\footnote{Suppose \textsc{Euclidean mixed model} is more than two-dimensional locally, then there are more than three atoms in a local neighborhood equidistant from each other. But their distances just are the Euclidean distances among their representative pairs of integers. Thus there are more then three pairs of integers that are equidistant from each other on the Euclidean plane. But this is known to be impossible. Thus, \textsc{\textsc{Euclidean mixed model}} is no more than two-dimensional locally. Moreover, it is clear that \textsc{Euclidean mixed model} is not one-dimensional locally, so it is exactly two-dimensional.} Although it is tricky to compare simplicity overall, this is one aspect that \textsc{Euclidean mixed model} may seem simpler.	As a bonus, under the mixed account, there is no need for defining dimensions to be relative to scales.\footnote{Forrest needs the definition of dimensionality to be relative to the scale because he wants to recover \textit{some} sense in which space is three (or four) dimensional.}  The dimension of a space can be uniformly determined by the local geometry:
	
	\begin{quote}
		\textsc{Dimension*.} A space is \textit{$N$-dimensional} iff $N$ is the least number such that every local neighborhood can be isometrically embedded in a $N$-dimensional continuous space.\footnote{This definition is analogous to the definition of the dimension of a manifold (i.e., a continuous space). One may try to translate this definition into a more intrinsic form such as this:
			\begin{quote}
				\textsc{Dimension$\dagger$.} A space is \textit{$N$-dimensional} iff $N$ is the least number that there are at most $N+1$ atoms that bear the same primitive distance to each other.  
			\end{quote}
			
			\noindent The problem with \textsc{Dimension$\dagger$} is that it leads to counterintuitive results. For instance, if no two pairs of atoms in the same local neighborhood have the same primitive distance, then \textsc{Dimension$\dagger$} would imply that the space is one-dimensional. But when such a space is not embeddable into one-dimensional continuous space, it is intuitively not one-dimensional.}
	\end{quote}

	In summary, although Forrest's account does make do with economical resources in one respect, having a full-fledged local metric (instead of just primitive adjacency) has important payoffs. Thus I recommend the mixed account as a response to Weyl's tile argument: it is versatile, compatible with our best physical theory and no less natural than the standard account of distance for continuous space, and it has distinctive advantages over Forrest's proposal.

	\newpage

	\begin{appendices}

		\section{}
		Now I shall turn to how well space approximates Euclidean space under the mixed account. Under this account, an atomistic space can be represented by a set of points with a \textit{shortest path metric} that assigns some pairs of points real-valued distances (bounded by a finite number) and derives other distances as their least sums. 
		
		We will understand ``approximation" in terms of ``almost isometry." Let $e(p,q)$ be the Euclidean distance between two points $p,q$ in Euclidean space. Let $\epsilon, r$ be two positive numbers. A metric space $X$ with a metric $d$ is \textit{$\epsilon$-isometric} to Euclidean space $E$ with regard to $r$ iff there is a map $f$ from $X$ to $E$ such that (1) for $x,y\in X$,  we have $$1-\epsilon\leq \frac{e(f(x),f(y))}{d(x, y)}\leq 1+\epsilon$$  (the smallest $\epsilon$ such that $f$ satisfies this condition is called the \textit{distortion} of $f$);\footnote{Here, the notion of approximation is cast in a different way from Forrest's (1995). Forrest showed that his model approximates Euclidean space in the sense that we can map Euclidean space into his model such that the distances are approximately preserved. Here, it is the other way around: a model approximates Euclidean space in the sense that we can map this model into Euclidean space that preserves distances approximately. I do not consider either interpretation of approximation to be better than the other, but I work with this one because I feel it a bit more natural.} (2) for every $p\in E$, there is a $x\in X$ such that $e(p,f(x))\leq r$. In other words, the embedded points cover $E$ reasonably well so that there are no obvious ``clusters" and ``holes." 
		
		\newtheorem{nonplanar}[subsection]{Theorem}
		
		\begin{nonplanar}
			For any $\epsilon$ and $r$, there is a set of points with a shortest path metric (with distances being bounded by a finite number) that is $\epsilon$-isometric to Euclidean space with regard to $r$.
		\end{nonplanar}
		
		\noindent \textit{Proof.} For brevity, I will resort to the following abbreviations when applicable. Given an embedding $f$ of a metric space into Euclidean space, for any points $x,y$ in the space, let $\|xy\|_f=e(f(x),f(y))$ (the subscript ``$f$" is omitted if it is clear which embedding we refer to). Also, for any points $p,q$ in Euclidean space, let $\|pq\|=e(p,q).$
		
		Let $G$ be an embedding of an infinite set $X$ to Euclidean space $E$ such that there is an $r$ such that for any $p\in E$, we can find an $x\in X$ with $e(p,G(x))<r$. (For example, if $G$ maps members of $X$ to Euclidean points represented by pairs of integers, then $r$ in question is at least $\sqrt{2}/2$.) We will construct a metric over $X$ such that the resulting metric space is $\epsilon$-isometric to Euclidean space under $G$, where $\epsilon$ is a small number we choose. 
		
		$M$ is a real-number parameter that will play an important role in assigning weights and in determining the distortion of the intended embedding. For any $x,y\in X$, if $\|xy\|> M$, we can find a sequence of points $p_1,p_2,...p_n$ in $E$ such that $p_0=G(x)$, $p_n=G(y)$, $\|p_0p_1\|=\|p_1p_2\|=...=\|p_{n-2}p_{n-1}\|=M$ and $\|p_{n-1}p_n\|<M$. Let $N=\|p_{n-1}p_n\|$. Consider $p_i,p_{i+1}$, where $i=1,...,n-2$. We can find $x_i,x_{i+1}\in X$ such that $e(G(x_i),p_i)<r$ and $e(G(x_{i+1}),p_{i+1})<r$. We know that the largest distance between points on two circles is equal to the distance between their centers plus their radii.\footnote{Here's a proof for the simple case in which two circles in question have the same radius, which is adequate for our purpose. Let two circles be $x_1=r\cos\theta_1$, $y_1=r\sin\theta_1$, $x_2=r\cos\theta_2+n$, $y_2=r\sin\theta_2.$ Then,  $(x_1-x_2)^2+(y_1-y_2)^2=n^2-2r^2\cos(\theta_1-\theta_2)-2nr(\cos\theta_1-\cos\theta_2)+2r^2\leq n^2+2r^2+4nr+2r^2=(n+2r)^2.$ That is, for two circles with the same size, the largest distance between two points on them is equal to the distance between their centers plus their radii.} Thus, $\|x_ix_{i+1}\|<M+2r$. Now, for any two $a,b\in X$, if $\|ab\|<M+2r$, then let the primitive distance $d(a,b)=\|ab\|$; otherwise, $d(a,b)$ is not defined. Then, $M\leq d(x_i,x_{i+1})<M+2r$. Moreover, it's easy to see that $M\leq d(x,x_1)\leq M+r$ and $N\leq d(x_{n-1},y)\leq N+r$. It follows that $d(x,y)\leq d(x,x_1)+d(x_1,x_2)+...+d(x_{n-1},y)< n\cdot (M+2r)+(N+r)$. Furthermore, if $x, x_1, ...x_n, y$ is a shortest path, then $d(x,y)=d(x,x_1)+d(x_1,x_2)+...+d(x_{n-1},y)=\|xx_1\|+....+\|x_{n-1}y\|\geq \|xy\|$. Thus, we have:
		
		$$1\leq \frac{d(x,y)}{\|xy\|}<\frac{n\cdot (M+2r)+(N+r)}{nM+N}=1+\frac{(2n+1)r}{nM+N}$$
		The distortion $\displaystyle\delta=\frac{d(x,y)}{\|xy\|}-1<\frac{(2n+1)r}{nM+N}<\frac{(2n+1)r}{nM}<\frac{3r}{M}$. Then, for any small positive number $\epsilon$, we can make $\delta<\epsilon$ by letting $M=3r/\epsilon$. (Note that if we are only concerned with distances that involve a large $n$, we only need $M$ to be $2r/\epsilon$.) This completes the case for any $x,y\in X$ with $\|xy\|> M$. If $\|xy\|\leq M$, then we have $d(x,y)=\|xy\|$, in which case there is no distortion. Therefore, we have found a metric space, in which all distances are bounded by $3r/\epsilon +2r$, that is $\epsilon$-isometric to Euclidean space at any scale. $\square$

	\end{appendices}

\end{document}